\documentclass[prl,twocolumn,showpacs,reprint,preprintnumbers,amsmath,amssymb]{revtex4}

\usepackage{bm}
\usepackage{graphicx}        % Figures
\usepackage{longtable}       % Tables
\usepackage{amsfonts}        % Fonts
\usepackage{amsmath}         % Includes useful equation modes
\usepackage{amssymb}         % Full set of symbol names included in the amsfonts
\usepackage{bm}              % Bold in math mode
\usepackage{color}           % For colored boxes
\usepackage{dcolumn}         % Align table columns on decimal point
\usepackage{epsfig}          % For eps figures
\usepackage{hyperref}        % Turn all your internal references into hyperlinks
\preprint{Accepted in Phys. Status Sol. A}

\begin{document}

% Title of the article
\title{Electronic properties of Cs-based halide perovskites: An \textit{ab-initio}
study}

% Authors
\author{Georgia Moschou}
  \author{Athanasios Koliogiorgos}
  \author{Iosif Galanakis}\email{galanakis@upatras.gr}

% author's affiliations/addresses
\affiliation{%
Department of Materials Science, School of Natural Sciences,
University of Patras,  GR-26504 Patra, Greece}

\date{\today}

\begin{abstract}
Halide perovskites consist a class of materials under intense investigation
due to their potential technological applications like solar cells, optoelectronic
devices and catalysis. Recently we have studied using electronic band
structure calculations from first principles, the cubic MABX$_3$ compounds [A. Koliogiorgos et al.,
Comput. Mater. Sci. \textbf{138},  92 (2017)], where MA
stands for the methylammonium cation, B is a divalent cation and X a halogen.
We expand our study in the case where Cs stands in place of the MA cation.
Our results suggest that the Cs-based compounds exhibit also a variety of lattice constants
and energy band gaps. The calculated equilibrium lattice constants differ substantially from the experimental
ones. The calculated energy gaps also show large deviations for these lattice constants.
Moreover, the use of more sophisticated functionals leads to conflicting changes in the energy gap values
and its effect is materials dependent.
Our results suggest that contrary to the MA halide perovskites, the Cs halide perovskites consist a
more delicate case and there is still a long way for \textit{ab-initio} calculations to accurate describe their structural
and electronic properties.
\end{abstract}

\maketitle   % please do not remove

\section{Introduction}\label{Intro}

Perovskites consist a wide class of materials embracing several hundreds of compounds,
since within the same chemical formula a wide range of components can be combined \cite{Chen2015,Gratzel2014}.
They were named after the Russian mineralogist Lev Perovski and the name was first used to describe the
CaTiO$_3$ compound \cite{Pena2001}.
Although several perovskite materials  exist in nature \cite{Science1}, the developments
in both chemistry and physics of materials led to the growth of novel perovskite materials suitable
for several technological applications like solar cells \cite{Science2}, optoelectronics \cite{Science3}
and catalysis \cite{Science4}.

Among perovskites, a very popular subclass are the ones having
the chemical formula ABX$_3$, where A is a monovalent cation, B a divalent cation
and X a halogen atom. Due to the presence of the halogen atom they are widely known as
halide perovskites \cite{Hoefler2017}. A wide range of chemical  elements
can be combined as long as the requirement for charge neutrality is
satisfied. Interestingly the A cation can be also a charged organic molecule, e.g.
methylammonium, formamidinium etc giving birth to the so-called hybrid or organic-inorganic
halide perovskites \cite{Gratzel2014,Hoefler2017,Yang2016,Papavassiliou2012}.
Among these hybrid halide perovskites special attention has been paid to the methylammonium (MA)
based ones \cite{Brittman2015,Frost2014,Filippetti2014,Albero2016,Zhou2016},
where the MA cation is CH$_3$NH$_3$, due to the fact that MAPbI$_3$ has an energy
gap of about 1.5-1.7 eV \cite{Zhao2015,Quarti2016} absorbing at the optical regime.

Motivated by the search for lead-free hybrid halide perovskites, in reference \cite{Koliogiorgos2017}
we carried out an extended \textit{ab-initio} study of the MABX$_3$
compounds crystallizing in the cubic structure. As X we considered all possible halogen atoms,
namely F, Cl, Br and I. As divalent B cations we took into account the
alkali earth elements (Ca, Sr, Ba) since they have two valence
\textit{p} electrons, the late transition metal atoms (Zn, Cd, Hg)
which have two valence \textit{s} electrons and the metalloids
(Ge, Sn, Pb) which have also two valence \textit{p} electrons but
contrary to the alkali earth elements the valence \textit{d}
states are completely occupied. As our results suggested in reference \cite{Koliogiorgos2017}
several of these 36 compounds possessed energy gaps within the optical regime,
making them suitable for solar cell applications.

\begin{figure}
\includegraphics[width=\columnwidth]{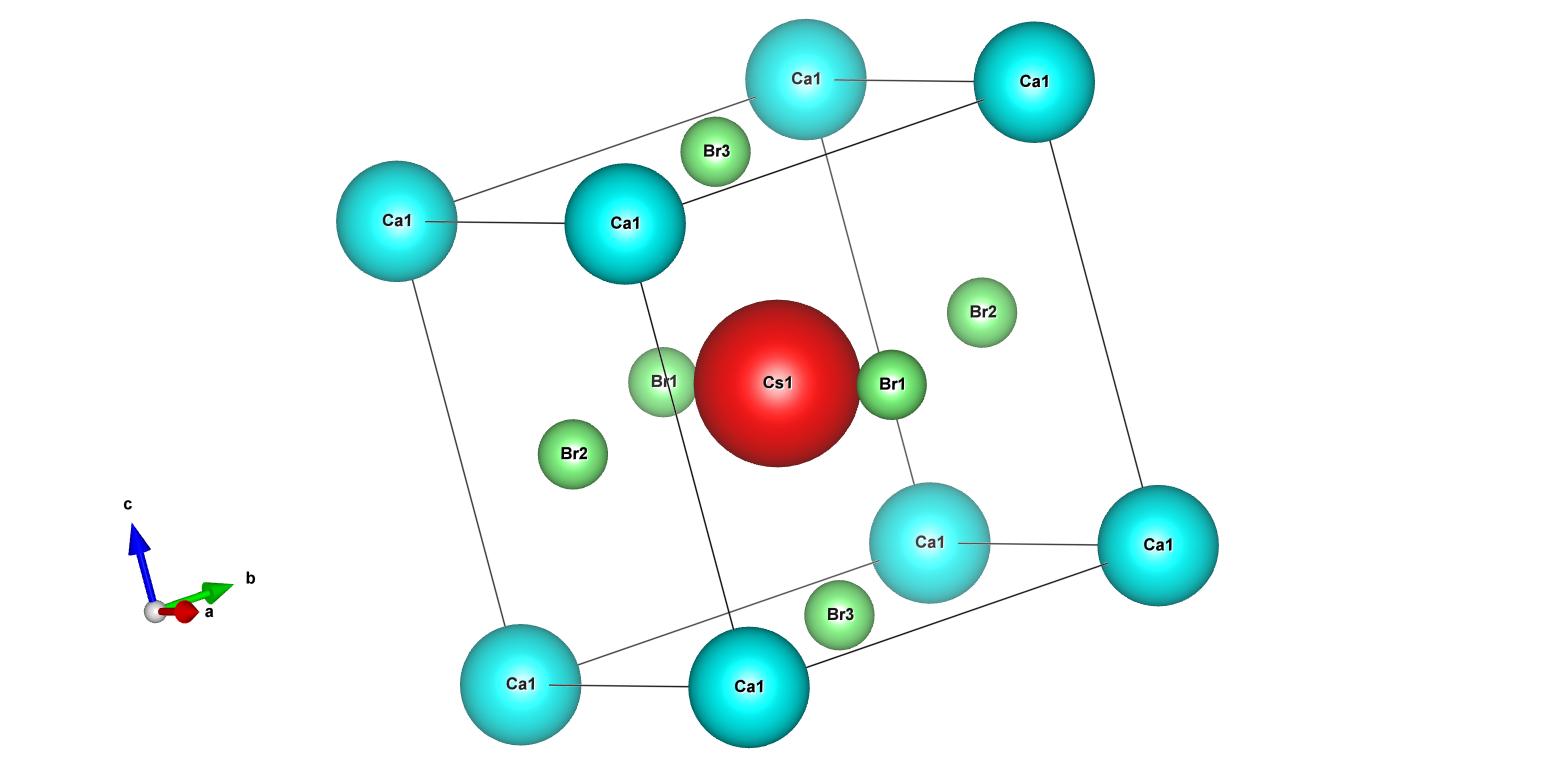}
\caption{Schematic representation of the cubic lattice structure assumed
for the CsBX$_3$ compounds. The Cs atom is at the center of the
cube, the divalent B cations are at the corners of the cube and
the halogen atoms are at the center of the faces.} \label{fig1}
\end{figure}

In the present study we expand our previous work in the case of the Cs-based compounds
having the chemical formula CsBX$_3$. First, we considered only the cubic structure, presented in figure \ref{fig1},
 taking into account all possible combination of divalent
cations and halogen anions as for the MABX$_3$ compounds mentioned above.
For these compounds we computed initially the equilibrium lattice constant and then their electronic
properties focusing on the width of the energy band gap. Second, we considered the experimental
lattice structures for the CsBX$_3$ which have been grown successfully and we have calculated
the energy band gap also in their case.  The latter crystallize not only in the cubic
but also in orthorhombic lattices, since as it is well known perovskite materials undergo several structural
transitions \cite{Hoefler2017}.
The manuscript is organized in the following way. In section \ref{comput}  we present the details of our
calculations. In section \ref{results} we present and discuss our results.
Finally, in section \ref{conclusions} we summarize our results and conclude.

\section{Computational method}\label{comput}

The \textit{ab-initio} electronic band structure which has been employed in
the present study is the so-called VASP (acronym for Vienna \textit{ab-initio}
Simulation Package) developed at
the Institut f\"ur Metaliphysik of the Universit\"at Wien
\cite{VASP}. As pseudopotentials we used the projector augmented waves (PAW)
\cite{PAW}. As exchange correlation functional we have used the so-called PBEsol
(Perdew-Burke-Ernzerhof for solids) \cite{PBE,PBEsol}, which belongs to the family
of the  generalized gradient approximation (GGA) functionals.
Although GGA functionals in general are known to reproduce accurately the
structural properties, they fail to reproduce the energy gap in semiconductors like
perovskites although they yield the correct shape of the bands.

Since the exact value of the energy gap is crucial for applications, we have also
used in our calculations the modified
Becke-Johnson functional in conjunction with the PBEsol one (known
as mBJ+PBEsol). The mBJ has been developed in 2006 by Becke and
Johnson in an attempt to provide an efficient exchange functional
which would reach the accuracy of the hybrid functionals like
Heyd-Scuseria-Ernzerhof (HSE06) but which would require CPU resources similar to the GGA
calculations \cite{Becke,Tran2009}. The mBJ functional is a potential-only functional
being a local approximation to an atomic exact-exchange potential
plus a screening term which is used in conjunction to one of the
exchange-correlation schemes (PBEsol in our case). Thus the
mBJ-based calculations within VASP are not self-consistent with
respect to the total energy \cite{manual-vasp}. Finally,  for the two compounds
which have been grown experimentally,
CsPbBr$_3$ and CsPbI$_3$, we have also performed calculations using the hybrid
HSE06 functional \cite{HSE06}, which has
been successfully implemented in VASP \cite{VASP-HSE06}. HSE06 is well-known to reproduce
accurate band gap values for semiconductors but it leads to very demanding calculations, both
in cpu time and cpu resources \cite{Franchini2014}. In the case of the MABX$_3$ compounds
studied in reference \cite{Koliogiorgos2017} both HSE06 and mBJ+PBEsol yielded similar
band gap values and thus we expect this to be the case also for the Cs-based compounds.
It is wort to notice, that GGA calculations for related ABO$_3$ perovskites underestimate the band gap value,
while the hybrid B3PW and B3LYP functionals allows to achieve much better agreement with the experiment for
the band gap values \cite{Eglitis1,Eglitis2}.

Concerning the details of the calculations, we have used for all
cubic compounds  in the first part of our study, a cutoff for the kinetic energy of the plane waves of 420
eV and for the Ge, Sn and Pb atoms we have included in the PAW
basis the 3\textit{d}, 4\textit{d} and 5\textit{d} orbitals,
respectively, as valence states. For the case of the mBJ+PBEsol
calculations we have used a more advanced basis set including also
the kinetic energy density of the core electrons. To determine the equilibrium lattice
constants we have used a 6$\times$6$\times$6
Monkhorst-Pack grid in the 1st Brillouin zone \cite{Monkhorst} in
conjunction with the PBEsol functional. For the equilibrium lattice constant, we have used a denser
10$\times$10$\times$10 grid to calculate the electronic properties.
For the compounds adopting the experimental lattice structure we present and discuss the details of the
calculations in section \ref{experiment} and in table \ref{table3}.

\begin{figure}[t]%
\includegraphics[width=\linewidth]{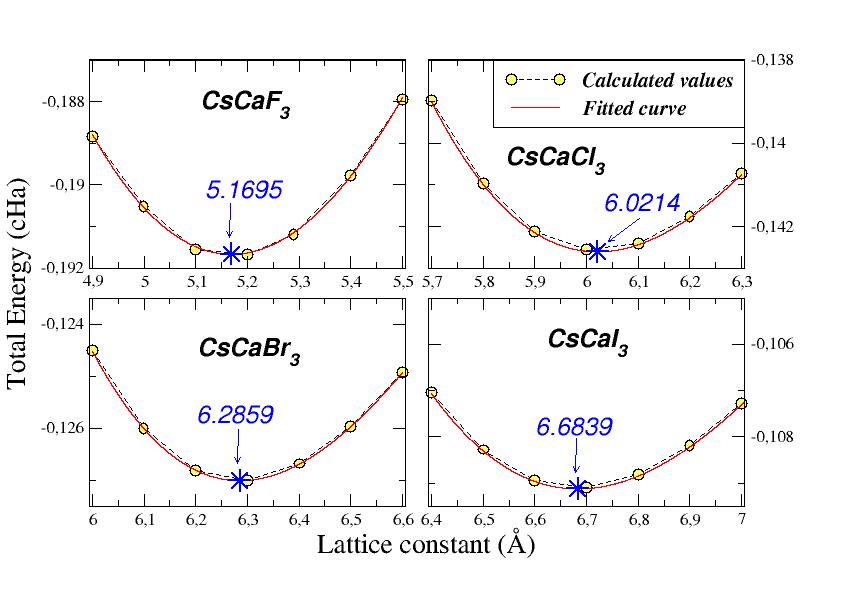}
\caption{%
 Calculated total energy versus the lattice constant for the cubic
CsCaX$_3$ compounds. The calculated values have been fitted with a
third order polynomial in order to determine the equilibrium
lattice constant. Note that the use of the Murnaghan equation of
state gives identical results in the case of the compounds under
study. } \label{fig2}
\end{figure}

\begin{table}
  \caption{Calculated equilibrium lattice constants, $a_{eq}$, in \AA\ for the studied cubic halide
  perovskites  using the PBEsol approximation. These perovskites have the chemical formula
  CsBX$_3$ where B is a divalent
  cation and X is a halogen atom.}
  \begin{tabular}{lllll}
    \hline
$a_{eq}^\mathrm{CsBX_3}$(\AA )&  X=F &  X=Cl&  X=Br & X=I \\
\hline

B=Ca &  5.1695 &   6.0214 &   6.2859 &  6.6839 \\
B=Sr &  5.1813 &   6.0313 &   6.2947 &  6.6881\\
B=Ba &  5.1931 &   6.0452 &   6.3089 &  6.7001\\
B=Zn &  5.0125 &   5.9734 &   6.3969 &  7.0772\\
B=Cd &  5.0608 &   6.0266 &   6.3710 &  6.9758\\
B=Hg &  4.9680 &   5.9715 &   6.5038 &  7.1538\\
B=Ge &  5.1153 &   6.0136 &   6.3015 &  6.7670\\
B=Sn &  5.1331 &   6.0216 &   6.3020 &  6.7178\\
B=Pb &  5.1614 &   6.0274 &   6.3008 &  6.7103\\

    \hline
  \end{tabular}
  \label{table1}
\end{table}

\begin{figure}[b]%
\includegraphics[width=\linewidth]{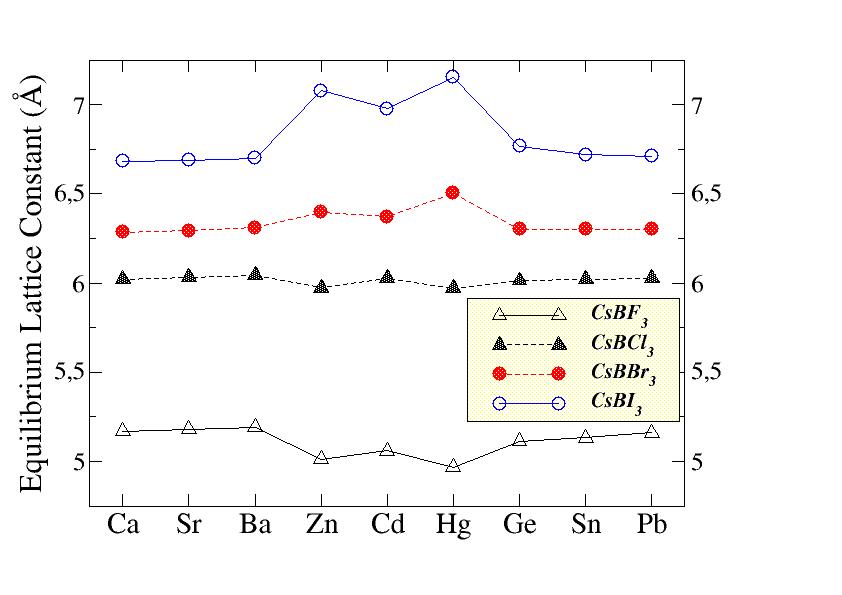}
\caption{%
 Behavior of the calculated equilibrium lattice constant for the cubic compounds in \AA\ as a function of the divalent
cation. Each line corresponds to a different halogen atom.}
\label{fig3}
\end{figure}

\section{Results and discussion}\label{results}

\subsection{Equilibrium lattice constants for the cubic compounds}\label{Structure}

As mentioned above, we have considered that the
compounds under study crystallize in the cubic
structure shown in figure \ref{fig1} \cite{Leguy2015}. The Cs atoms sit at
the center of the cube, the
divalent cations sit at the corners of the cube surrounded by
halogen atoms in an octahedral environment. To determine the equilibrium
lattice constant we have calculated the total energy for seven lattice constant values
around the minimum. Then we have fitted a third order polynomial curve to determine
the equilibrium lattice constant corresponding to the minimum of
the energy. In figure \ref{fig2} we present the calculated total
energy values versus the lattice constant and the fitted
curve for the four CsCaX$_3$ compounds. In all cases the fitting is extremely
good passing from all seven calculated points.

In table \ref{table1}, we have gathered the calculated equilibrium lattice constants in
\AA\ for all 36 calculated compounds and in figure \ref{fig3} we have plotted them
as a function of the divalent cation to make  their behavior clear.
The obtained equilibrium lattice constants scan a
wide range of values starting from 4.9680 \AA\ in the case of
CsHgF$_3$ up to 7.1538 \AA\ for CsHgI$_3$. All calculated values are larger
than the values for the corresponding MA compounds studied in reference
\cite{Koliogiorgos2017} due to the larger atomic radius of Cs compared to the MA cation.
In figure \ref{fig3} it is easier to trace the trends in the equilibrium lattice constants.
When the halogen atom is Cl almost all compounds under study have the same lattice constant.
In the case of the other three halogens, the compounds where the divalent cation is
an alkali earth metal or a metalloid have comparable values of equilibrium lattice constants.
When the divalent cations is one of the three late transition metal atoms, Zn, Cd or Hg,
the equilibrium lattice constants show a different behavior and the trend is opposite when
the halogen atom is F or I. In the former case the Zn, Cd and Hg based compounds have smaller
lattice constant than the other divalent compounds, while in the case of the iodides the Zn,
 Cd and Hg have considerably larger equilibrium lattice constants with respect to the other divalent cations.

\begin{table}
  \caption{Calculated energy gap in eV  using the
mBJ+PBEsol and PBEsol functionals (the latter in parentheses)
for the cubic CsBX$_3$ halide perovskites and for the equilibrium lattice in table \ref{table1}.
The zero values correspond to
a gapless (zero-gap) semiconducting behavior. We also provide the
HSE06 values in braces where available.}
  \begin{tabular}{lllll}
    \hline
& \multicolumn{4}{c}{Band gap (eV)} \\
& \multicolumn{4}{c}{mBJ+PBEsol (PBEsol) [HSE06]} \\
CsBX$_3$  &  X=F &  X=Cl&  X=Br & X=I \\
\hline

B=Ca&1.62(0.59)  &  0.94(0.58) &   0.95(0.41)  &  0.60(0.24)  \\
B=Sr&2.51(1.40)  &  1.87(1.10) &   1.59(0.79)  &  1.29(0.64)  \\
B=Ba&3.30(2.14)  &  2.66(1.88) &   2.40(1.50)  &  1.26(1.26)  \\
B=Zn&0.46(0.00)  &  0.13(0.00) &   0.27(0.00)  &  0.36(0.00)  \\
B=Cd&0.21(0.00)  &  0.00(0.00) &   0.09(0.00)  &  0.09(0.00)  \\
B=Hg&0.00(0.00)  &  0.00(0.00) &   0.00(0.00)  &  0.00(0.00)  \\
B=Ge&0.79(0.00)  &  0.40(0.00) &   0.52(0.00)  &  0.28(0.00)  \\
B=Sn&0.75(0.00)  &  0.47(0.00) &   0.46(0.00)  &  0.45(0.00)  \\
B=Pb&0.82(0.00)  &  0.45(0.00) &   0.53(0.00)[0.49]  &  0.33(0.00)[0.38]  \\
    \hline
  \end{tabular}
  \label{table2}
\end{table}

\subsection{Electronic and gap properties at the theoretical equilibrium lattice
constants}\label{gap}

Since we have determined the equilibrium lattice constants, we
proceeded with the calculation of the energy gaps which is also
the main finding of the present study. First, we employed the PBEsol
functional. As mentioned above, the former is not accurate enough
in most cases to compute the energy gaps, and thus we used the
PBEsol calculated electronic charge and wavefunctions as the input
to perform electronic band structure calculations with the most
accurate mBJ+PBEsol using
the same grid in the reciprocal space as for the PBEsol
calculations. In table \ref{table2} we have gathered both the mBJ+PBEsol and PBEsol
(the latter in parenthesis)
calculated energy gaps for all 36  compounds under study.
When the divalent cation was a metalloid or a late transition-metal atom,
PBEsol led to gapless semiconductors, \textit{i.e.} the gap is vanishing
being almost zero. In the case of the alkali earth metals PBEsol produced sizeable gaps
ranging from 0.34 eV for CsCaI$_3$ to 2.14 eV for CsBaF$_3$.

\begin{figure}[t]%
\includegraphics[width=\linewidth]{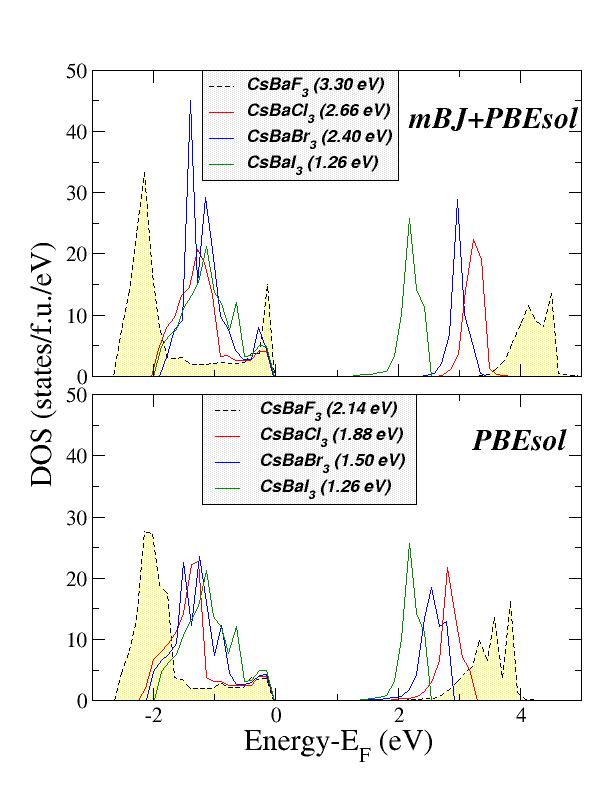}
\caption{%
DOS per formula unit (f.u.) as a function of
the energy for the cubic CsBa(F, Cl, Br, I)$_3$ compounds for the theoretical
equilibrium lattice constants using the
PBEsol (lower panel) and MBJ+PBEsol (upper panel) functionals. The
zero energy corresponds to the Fermi level. Note that the use of
the mBJ+PBEsol functional opens the gap but does not influence
the character of the bands. } \label{fig4}
\end{figure}

The use of mBJ+PBEsol opened the gaps in all cases with the exception of the
Hg-based compounds.  In general the present compounds present very small
values of the energy band gaps with respect to the corresponding MA compounds in reference
\cite{Koliogiorgos2017}. With few exceptions where the gap is almost conserved,
the energy band gap becomes smaller as we change the halogen atom going
from the light F to the heavy I one. This is
expected since the lighter the halogen atom is, the deeper are its
valence \textit{p} states and the larger is the energy gap,
\textit{i.e.}, in the case of F the valence states are the
2\textit{p} orbitals while in the case of I the valence states are
the 5\textit{p} orbitals which are much higher in energy.
Some of the compounds containing alkali earth metals, CsCaF$_3$,
CsSrCl$_3$ and CsSrBr$_3$ have band gaps within the optical regime. In order to elaborate
more on the effect of the different functionals, we present in figure \ref{fig4} the density
of states (DOS) for the CsBaX$_3$ compounds. The mBJ+PBEsol functional, first, shifts the
conduction band higher in energy opening the gap with respect to the PBEsol functional.
Second, it slightly narrows the bands leading to peaks of larger intensity.

\begin{figure}[t]%
\includegraphics[width=\linewidth]{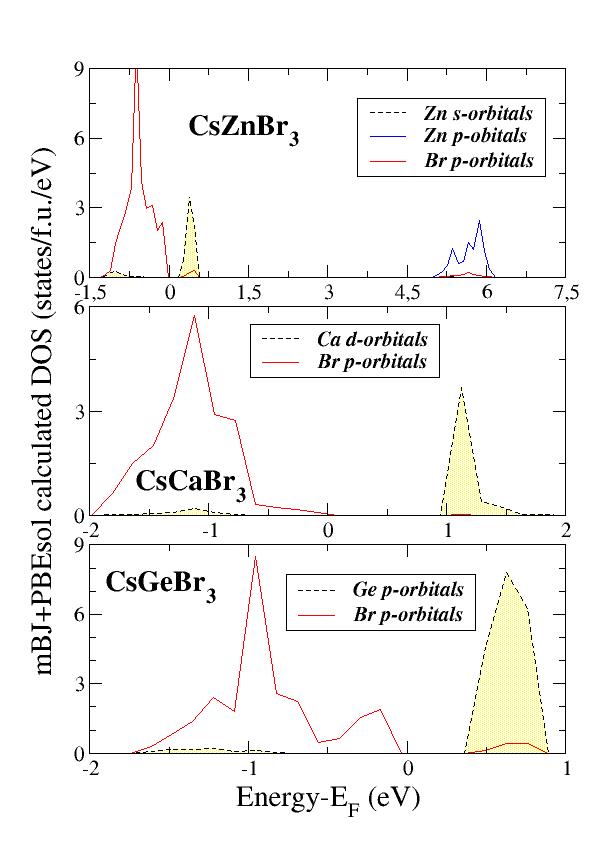}
\caption{%
Atom- and orbital resolved DOS as a function
of the energy for the  CsZnBr$_3$ (upper panel), CsCaBr$_3$
(middle panel) and CsGeBr$_3$ (lower panel) compounds and for the
theoretical equilibrium lattice constant using the
MBJ+PBEsol functional. The valence bands for all three compounds
consist of the halogen \emph{p}-orbitals. The conduction bands are
made up from the Zn \emph{s}-, Ca \emph{d}- and Ge
\emph{p}-states, respectively. The kind of hybridization
determines the width of the energy gap.  } \label{fig5}
\end{figure}

To elucidate the trends observed in the calculated band gap
energies,  we have plotted in figure \ref{fig5} the atom- and
orbital resolved DOS for the
Cs(Zn, Ca or Ge)Br$_3$ compounds using the mBJ+PBEsol functional.
In all three cases the valence bands
are made up from the \textit{p} states of the Br atoms. Note that in
the cubic structure  all three Br atoms are equivalent as
shown in figure \ref{fig1}. There
is also a very small \textit{s} admixture which is not shown here.
 In the case of CsGeBr$_3$ the conduction band is
made up from the Ge empty 4\textit{p} states and thus the gap is
due to the \textit{p}-\textit{p} hybridization; a similar situation
occurs if instead of Ge we have the isovalent Sn or Pb.
In the case of Ca, and the isovalent Sr and Ba atoms, the conduction band
is now made up from the unoccupied  triple degenerate $t_{2g}$
\textit{d}-orbitals which transform
in the same way as the \textit{p} orbitals in the case of
tetrahedral and octahedral symmetries. The empty \textit{p}-states
are higher in energy and are not shown in the figure.
Finally, when the cation is one of the transition metal atoms (Zn, Cd or
Hg), a conduction \textit{s}-state appears clearly between the
occupied \textit{p} states of the halogen atoms and the unoccupied
\textit{p} states of the Zn, Cd or Hg atoms. Thus although the
\textit{p}-\textit{p} hybridization opens a sizeable gap, the
location of the Zn, Cd, or Hg \textit{s}-states within the gap
leads to much smaller energy gap values and even to the observed
gapless  behavior shown in table \ref{table2}.

\begin{figure}[t]%
\includegraphics[width=\linewidth]{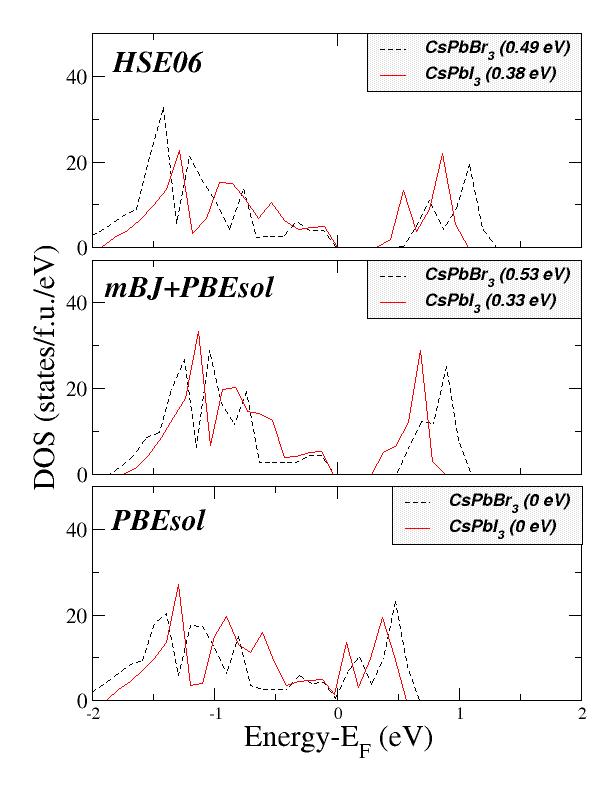}
\caption{%
DOS per f.u. as a function of
the energy for the cubic CsPbBr$_3$  and CsPbI$_3$ compounds at their
theoretical equilibrium lattice constant using the
PBEsol (lower panel), mBJ+PBEsol (middle panel) and HSE06 (upper
panel) functionals. HSE06 and mBJ+PBEsol open the gap with respect
to PBEsol but HSE06 keeps the form of the bands while mBJ+PBEsol
leads to narrower bands. } \label{fig6}
\end{figure}

For the CsPbBr$_3$ and CsPbI$_3$ compounds we performed  calculations using also the HSE06 functional
which is expected to be the state-of-the-art for the computation of
energy band gaps in semiconductors to compare it with the mBJ+PBEsol values. HSE06 gave values very close to the
mBJ+PBEsol functional; 0.49 eV and 0.38 eV for the CsPbBr$_3$ and CsPbI$_3$ compounds, respectively.
This behavior is similar to the MA compounds studied in reference \cite{Koliogiorgos2017}.
In figure \ref{fig6} we have plotted for both compounds the DOS
using all three PBEsol, mBJ+PBEsol and HSE06 functionals.
As can be deduced from the figure HSE06 shifts the conduction bands to
higher energy values, similarly to the mBJ+PBEsol functional, but contrary to the latter it
does not affect the shape of the bands.

\begin{table*}
  \caption{For the compounds which have been grown experimentally, we present
  the reference,, the details of the experimental lattice structure, the \textbf{k}-grid used
  in the reciprocal space and the calculated energy band gaps in eV using the
PBEsol and mBJ+PBEsol functionals.}
  \begin{tabular}{lllllll}
    \hline
Compounds    &  Space group &Lattice & Lattice constants (\AA ) & \textbf{k}-point grid   & PBEsol (eV) & mBJ+PBEsol (eV) \\ \hline
CsSnCl$_3$\cite{Bulanova1972} &Pm3m &cubic          &5.504              & 10x10x10   & 0.92&    0.97 \\
CsSnI$_3$ \cite{Chung2012} &Pnma &orthorhombic   &10.349x4.763x17.684& 4x8x2      & 1.89&    0.96  \\
CsPbF$_3$ \cite{Berastegui2001} &Pm3m & cubic         &4.475              & 10x10x10   & 2.62&    3.94\\
CsPbCl$_3$ \cite{Hutton1979} &Pm3m &cubic          &5.605              & 10x10x10   & 1.94&    2.04  \\
CsPbBr$_3$ \cite{Sakata1979}&Pm3m &cubic          &5.605              & 10x10x10   & 1.12&    0.69 \\
CsPbBr$_3$ \cite{Rodova2003}&Pm3m &cubic          &5.870              & 10x10x10   & 1.72&    0.98 \\
CsPbBr$_3$ \cite{Stoumpos2013}&Pnma &orthorhombic   &8.244x11.735x8.198 & 6x4x6      & 1.61&    0.78 \\
CsPbBr$_3$ \cite{Rodova2003}&Pbnm &orthorhombic   &8.207x8.255x11.759 & 6x6x4      & 1.61&    0.88 \\
CsPbI$_3$ \cite{Eperon2015}  &Pm3m &cubic          &6.1769             & 10x10x10   & 1.40&    0.67 \\
CsPbI$_3$  \cite{Stoumpos2013b}&Pnma &orthorhombic   &10.434x4.791x17.761& 4x7x2      & 2.29&    1.07 \\
    \hline
  \end{tabular}
  \label{table3}
\end{table*}

\subsection{Electronic properties for the experimental lattice structures}\label{experiment}

The energy band gaps which were calculated in the above section deviate strongly from the experimentally
determined bands gaps for the cubic CsPbI$_3$ in reference \cite{Eperon2015}. The theoretical value in table \ref{table2}
is 0.33 eV using the mBJ+PBEsol functional while the experimental one is 1.73 eV \cite{Eperon2015}. A close look in reference
\cite{Eperon2015} reveals that the experimental lattice constant of 6.1769 \AA\ deviates also strongly from the theoretically
determined equilibrium lattice constant of 6.7103 \AA . Thus one could assume that the discrepancy stems from
the difference in the lattice constants. To elucidate more we have carried out \textit{ab-initio}
calculations for a series of Cs-based halide
perovskites which have been grown experimentally. We have gathered in table \ref{table3} the reference
where the compounds have been grown, the details of the lattice structure, the details of the calculations and the calculated band gaps
using both PBEsol and mBJ+PBEsol functionals. The lattice of these compounds as determined experimentally are either
cubic or orthorhombic. For the orthorhombic the energy cutoff used is 280 eV. The compounds crystallizing in the cubic
structure of figure \ref{fig1} are CsSnCl$_3$ \cite{Bulanova1972},  CsPbCl$_3$ \cite{Hutton1979}, CsPbBr$_3$
\cite{Sakata1979,Rodova2003} and CsPbI$_3$ \cite{Eperon2015}; note that for CsPbBr$_3$ there are two different experimental
lattice constants of 5.605 \AA\ \cite{Sakata1979} and 5.870 \AA\ \cite{Rodova2003}.
CsPbF$_3$ crystallizes also in a cubic structure but now the
F atoms are located at the center of the edges and not at the center of the faces \cite{Berastegui2001}.
If we compare the experimental lattice constants for the cubic compounds in table \ref{table3} with the
theoretical equilibrium lattice constants in table \ref{table1}, we can remark that the deviations are very important; for example for CsSnCl$_3$
the experimental lattice constant is 5.504 \AA\ and the theoretical one 6.0216 \AA . In all cases the experimental
lattice constants are about 0.5 \AA\ smaller than the theoretical ones. Thus contrary to the MA-based halide perovskites,
in the case of the Cs-based halide perovskites, PBEsol fails to reproduce accurately the lattice constants.

The smaller lattice constants lead also to large deviations in the calculated band gaps. Contrary to the theoretical equilibrium
lattice constants, where PBEsol produced gapless behavior for all compounds where the divalent cation was Sn or Pb, in the
case of the  cubic experimental lattice constants the  PBEsol energy band gaps are sizeable and range from 0.97 eV for CsSnCl$_3$ to 1.72 eV
for cubic CsPbBr$_3$ where the lattice constant is 5.870 \AA\ \cite{Rodova2003}. Interestingly when the lattice constant of
CsPbBr$_3$ drops to 5.605 \AA\ \cite{Sakata1979}, the energy band gap shrinks to 1.12 eV. The calculated energy band gap for CsPbI$_3$
using PBEsol is 1.40 eV close to the experimental value of 1.73 eV \cite{Eperon2015}. The PBEsol calculated band gaps for the orthorhombic
structures are slightly larger reaching the 2.29 eV for CsPbI$_3$ \cite{Stoumpos2013b}. We can compare our calculated value of
1.61 eV for CsPbBr$_3$ in the orthorhombic structure with the experimental value of 2.25 eV \cite{Stoumpos2013}. We can see that the discrepancy
is considerably large. Thus we employed for all compounds under study also the mBJ+PBEsol functional. As discussed in detail in reference
\cite{Koliogiorgos2017}, mBJ+PBEsol is not guaranteed to open the energy band gaps with respect to PBEsol and its
effect is material specific. This is also reflected in our results. There are compounds like CsPbF$_3$ where mBJ+PBEsol increases the band gap
from 2.62 eV to 3.94 eV as was the case for the totality of MA-halide perovskites in reference \cite{Koliogiorgos2017}, but
in most cases presented in table \ref{table3} mBJ+PBEsol leads to a shrinking of the compounds and cubic CsPbI$_3$ shows a
gap of 0.67 eV and orthorhombic CsPbBr$_3$ of 0.78 eV deteriorating the agreement between calculations and experiments
\cite{Eperon2015,Stoumpos2013}. We were not able to perform further calculations using the HSE06 functional due to our limitations
in computer resources.

\section{Summary and conclusions}\label{conclusions}

Perovskites are materials of particular technological interest
due to their versatile applications ranging from solar cells to
optoelctronic devices and catalysis.
In an attempt to identify halide perovskites, we employed
\textit{ab-initio} electronic structure calculations and studied, first, the structural
and electronic properties of cubic CsBX$_3$ compounds with the divalent cation B being
one of the Ca, Sr, Ba, Zn, Cd, Hg, Ge, Sn or Pb atoms
considering all possible halogen atoms  X= F, Cl, Br and I.  The
first step in our study was to use the VASP \textit{ab-initio}
technique in conjunction with the Perdew-Burke-Ernzerhof for solids (PBEsol) functional for
the exchange-correlation to determine the equilibrium lattice
constants for all 36 compounds. Then for the theoretical equilibrium lattice constant
we made use of the more sophisticated mBJ+PBEsol functional and
computed the energy band gaps. In most cases usual PBEsol produced
gapless semiconductors, but mBJ+PBEsol led to the appearance of energy band
gaps.

Since the calculated equilibrium lattice constants differed substantially
form the experimental ones (where available), we also carried out electronic
structure calculations for several compounds which have been grown experimentally
and for which data existed on their lattice structure. These compounds crystallize both in the cubic as well
as the orthorhombic structure. The energy gaps calculated with PBEsol were greatly enhanced with respect to the
ones at the calculated equilibrium lattice constant. But in most cases the hybrid modified Becke-Johnson (mBJ)+PBEsol
functionals led to narrower energy band gaps.

Our calculations suggest that the case of CsBX$_3$ halide perovskites, unlike other
families of halide perovskites, is not trivial and discrepancies between
theory and experiments as well as between various functionals occur.
Thus further investigation is needed to clarify these points and allow
for a more accurate study of these materials in the future.

\section*{Acknowledgements}
Authors acknowledge financial support from the  project PERMASOL
(FFG project number: 848929).

\end{document}